\documentclass{article}

\PassOptionsToPackage{numbers, compress}{natbib}
\usepackage[final]{ml4mols_2020}
\usepackage[utf8]{inputenc} 
\usepackage[T1]{fontenc}    
\usepackage{hyperref}       
\usepackage{url}            
\usepackage{booktabs}       
\usepackage{amsfonts}       
\usepackage{nicefrac}       
\usepackage{microtype}      
\usepackage{amsmath}
\usepackage{physics}
\usepackage{subcaption}
\usepackage{chemformula}
\usepackage{graphicx}
\usepackage{color}
\title{A Kernel based Machine Learning Approach to Computing Quasiparticle Energies within Many-Body Green’s Functions Theory}

\author{%
  Gianluca Tirimb\`o\\
  Eindhoven University of Technology\\
  \texttt{g.tirimbo@tue.nl} \\
   \And
   Onur \c Caylak \\
   Eindhoven University of Technology \\
   \texttt{o.caylak@tue.nl} \\
   \AND
   Bj\"orn Baumeier \\
   Eindhoven University of Technology \\
   \texttt{b.baumeier@tue.nl} \\
}

\begin{document}

\maketitle

\begin{abstract}
We present a Kernel Ridge Regression (KRR) based supervised learning method combined with Genetic Algorithms (GAs) for the calculation of quasiparticle energies within Many-Body Green's Functions Theory. These energies representing electronic excitations of a material are solutions to a set of non-linear equations, containing the electron self-energy (SE) in the $GW$ approximation. Due to the frequency-dependence of this SE, standard approaches are computationally expensive and may yield non-physical solutions, in particular for larger systems. In our proposed model, we use KRR as a self-adaptive surrogate model which reduces the number of explicit calculations of the SE. Transforming the standard fixed-point problem of finding quasiparticle energies into a global optimization problem with a suitably defined fitness function, application of the GA yields uniquely the physically relevant solution. We demonstrate the applicability of our method for a set of molecules from the $GW$100 dataset, which are known to exhibit a particularly problematic structure of the SE. Results of the KRR-GA model agree within less than 0.01 eV with the reference standard implementation, while reducing the number of required SE evaluations roughly by a factor of ten.
\end{abstract}

\section{Introduction}
Quasiparticle (QP) excitations are a generalization of the concept of single-particle orbitals to interacting electron systems. Precise knowledge of the QP energy is crucial to understand direct and inverse photoemission~\cite{Lischner2013PhysicalStudy,Gruneis2008Electron-electronStudy,Tirimbo2020QuantitativeEmbedding}, tunneling~\cite{Dial2012ObservationsSpectroscopy}, or transport experiments, and vital for the calculation of optical absorption and reflectivity spectra~\cite{Rohlfing1998Electron-holeInsulators,Blase2020TheChemistry,Golze2019TheSpectroscopy}. Here, we focus on the $GW$ approximation of Many-Body Green's Function Theory~\cite{Sham1966Many-particleExciton,Hedin1970EffectsSolids}. It describes charged excitations, i.e when electrons are added/removed from a system of $N$ electrons. The key ingredient of the $GW$ framework is the evaluation of the electron's self-energy $\Sigma(\omega)$, the non-Hermitian operator which contains all many-body exchange and correlation effects beyond the electrostatic Hartree potential. 

QP energies depend self-consistently on $\Sigma(\omega)$. In general, the self-energy can be highly structured in frequency, leading to multiple QP solutions. The usual way to consider them all is to identify a frequency interval containing all potential solutions, evaluating $\Sigma(\omega)$ on a grid inside this interval, and then refining this using a bisection method (graphical solution). It is often not clear a priori how to choose the range of the interval and an appropriate grid spacing, and how to select regions of importance. Such a brute-force approach results in a large number of self-energy evaluations, with prohibitively high computational demand for larger systems. It is desirable to limit this number by exploiting a surrogate machine learning model. 

The application of machine learning and evolutionary computation models to quantum chemical calculations has been growing in popularity over recent years~\cite{ vonLilienfeld2020ExploringLearning, Bartok2017MachineMolecules, Gomez-Bombarelli2018AutomaticMolecules, Montavon2012LearningPrediction}. Neural networks, genetic algorithms, and kernel-based models have become popular approaches for generating surrogate property models allowing for fast predictions of relevant quantum molecular properties $P$~\cite{Caylak2020WassersteinRepresentations,Caylak2019EvolutionaryMaterials,Behler2015ConstructingReview,Behler2016Perspective:Simulations,Ceriotti2019UnsupervisedUnderstanding,Rupp2018GuestChemistry}. Specifically, kernel ridge regression (KRR) models approximate such properties with 
\begin{equation}
    \hat{P}(\mathbf{x}) = \sum_i^{M} \beta_i k(\text{dist}(\mathbf{x},\mathbf{x}_i)),
\end{equation}
where $M$ is the number of training samples, $\{\beta_i\}$ are the regression coefficients, $k$ is the kernel, and $\text{dist}(\mathbf{x},\mathbf{x}_i)$ is the distance measure between training samples $\{\mathbf{x}_i\}$ and $\mathbf{x}$. We focus on the incorporation of KRR models into the $GW$ framework to drastically reduce the computationally heavy self-energy evaluations.

\begin{figure}[t]
  \centering
  \includegraphics[width=\linewidth]{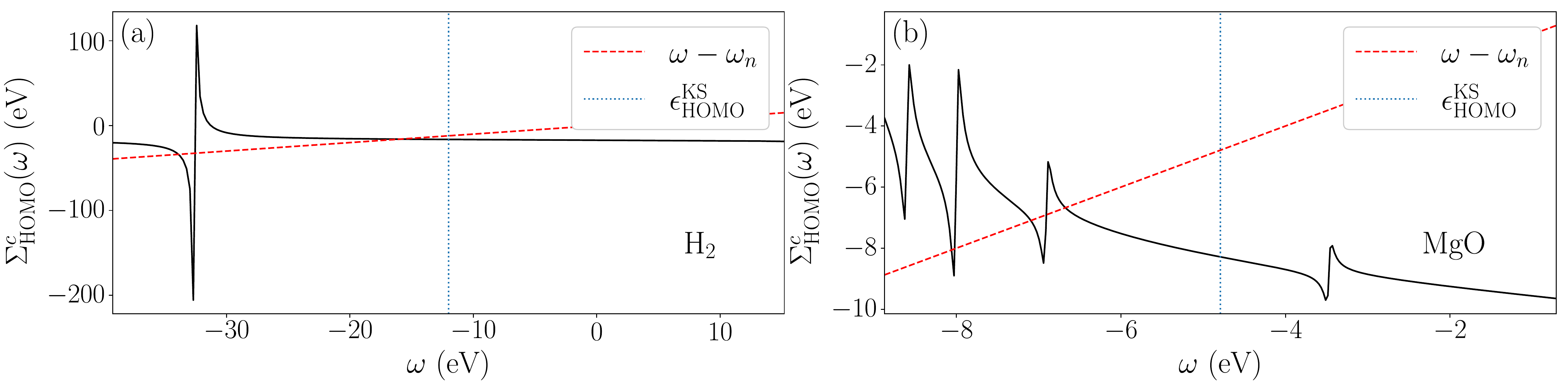}
  \caption{Graphical solution of Eq.~(\ref{equ:theory:qpequgw}) for the HOMO level of (a) \ch{H2} and (b) \ch{MgO}. Data generated with VOTCA-XTP~\cite{Wehner2018ElectronicVOTCA-XTP,Tirimbo2020Excited-stateVOTCA-XTP} code for G0W0@PBE~\cite{Perdew1996GeneralizedSimple} with def2-QZVP basis set~\cite{Weigend2005BalancedAccuracy}.
  }
  \label{fig:selfenergy}
\end{figure}

\section{Theoretical Background}
Quasiparticle energies $\varepsilon^{\mathrm{QP}}_n$ are the poles of the interacting one-electron Green's function, $G(\mathbf{r},\mathbf{r}',\omega)$~\cite{Sham1966Many-particleExciton,Hedin1970EffectsSolids}, and can be calculated by solving the quasiparticle or Dyson's equation
\begin{equation}
\label{equ:qpequation}
 [{H}_{0} + {\Sigma}(\epsilon^\mathrm{QP}_n) ]\ket{\psi^\mathrm{QP}_n} = \varepsilon^\mathrm{QP}_n \ket{\psi^\mathrm{QP}_n} ,
\end{equation}
where $H_0=-\frac{\nabla^2}{2}+V_{\mathrm{nuc}}+V_{\mathrm{Hartree}}$ is the effective single-body electronic Hamiltonian. $V_{\mathrm{Hartree}}$ ($V_{\mathrm{nuc}}$) is the classical electrostatic potential formed by the electrons (nuclei).  $\Sigma(\omega)$ is the electron self-energy operator that describes the exchange-correlation many-body effects. Kohn-Sham Density Functional Theory~\cite{Hohenberg1964InhomogeneousGas,Kohn1965Self-consistentEffects,Onida2002ElectronicApproaches} (KS-DFT) is an approximation to Eq.~(\ref{equ:qpequation}), in which the self-energy is given by the exchange-correlation potential, $\Sigma \sim V_\mathrm{xc}$.  The $GW$ approximation goes beyond KS-DFT and is an approximate solution of a set of equations known as Hedin's equations~\cite{Hedin1970EffectsSolids}, where the self-energy is expressed as 
\begin{equation}
\label{equ:GWselfenergy}
    {\Sigma}(\mathbf{r},\mathbf{r'},\omega) = \frac{i}{2 \pi} \int^{\infty}_{-\infty} {G}(\mathbf{r},\mathbf{r'},\omega + \omega') {W}(\mathbf{r},\mathbf{r'},\omega') d\omega' .
\end{equation}
Here, $W(\mathbf{r},\mathbf{r}',\omega')$ is the screened Coulomb interaction between unit charges at position $\mathbf{r}$ and $\mathbf{r'}$ calculated in the Random-Phase Approximation~\cite{Hybertsen1985First-principlesInsulators}. The replacement of the bare Coulomb interaction accounts for the interaction of an electron (or a hole) with the polarization response it induces in the medium. In the $G_0W_0$ approximation to  Eq.~(\ref{equ:GWselfenergy}) both $G$ and $W$ are constructed based on KS energies and wavefunctions. Perturbative QP solutions are obtained expanding Eq.~(\ref{equ:qpequation}) in a basis of KS states and neglecting off-diagonal elements of the self-energy operator. In this scenario, QP energies satisfy the following fixed-point problem
\begin{equation}
\label{equ:theory:qpequgw}
     \omega - \omega_{n} = \Sigma^\mathrm{c}_n(\omega)
\end{equation}
with $\omega_{n} = \varepsilon^\mathrm{KS}_{n}  + \bra{\psi^\mathrm{KS}_n}\Sigma^\mathrm{x} - V_\mathrm{xc}\ket{\psi^\mathrm{KS}_n}$ and $\bra{\psi^\mathrm{KS}_n} \Sigma^\mathrm{c}(\omega) \ket{\psi^\mathrm{KS}_n} = \Sigma^\mathrm{c}_n(\omega)$. We have split the self-energy into a frequency-independent exchange term $\Sigma^\mathrm{x}$ and a correlation part $\Sigma^\mathrm{c}(\omega)$, whose computation can be challenging since it involves summation over all combinations of occupied and empty states to determine the polarizability matrix and the calculation of its inverse for each frequency. Due to the pole structure of the self-energy as illustrated in Fig.~\ref{fig:selfenergy} for \ch{H2} and \ch{MgO}, respectively, there are in general several solutions to Eq.~(\ref{equ:theory:qpequgw}). The spectral weight, 
\begin{equation}
\label{equ:spectralweight}
Z_n(\omega) = \left(1-\frac{d \Sigma_n^\mathrm{c}(\omega)}{d \omega}\right) ^{-1},
\end{equation}
is used to identify the sought QP energy by $Z_n(\omega) \approx 1$, or $\left\vert {d \Sigma_n^\mathrm{c}(\omega)}/{d \omega} \right\vert \approx 0$. 

For \ch{H2}, two candidate solutions for the QP energy are found. From the spectral weight, it is clear that the actual solution lies in the frequency range in which the self-energy has no poles and is monotonously decreasing. In contrast, the situation is less straightforward for the highest occupied molecular orbital (HOMO) of \ch{MgO}, for which we can identify several potential QP energies with similar weights in the more structured parts of the self-energy. That these details cannot be known beforehand emphasizes that the reliable determination of the QP energy relies in general on the knowledge of $\Sigma^\mathrm{c}_n(\omega)$ and its derivative at a fine resolution over a wider frequency interval. 

\section{Incorporation of an active kernel machine learning model}

\begin{figure}[t]
  \centering
  \includegraphics[width=\linewidth]{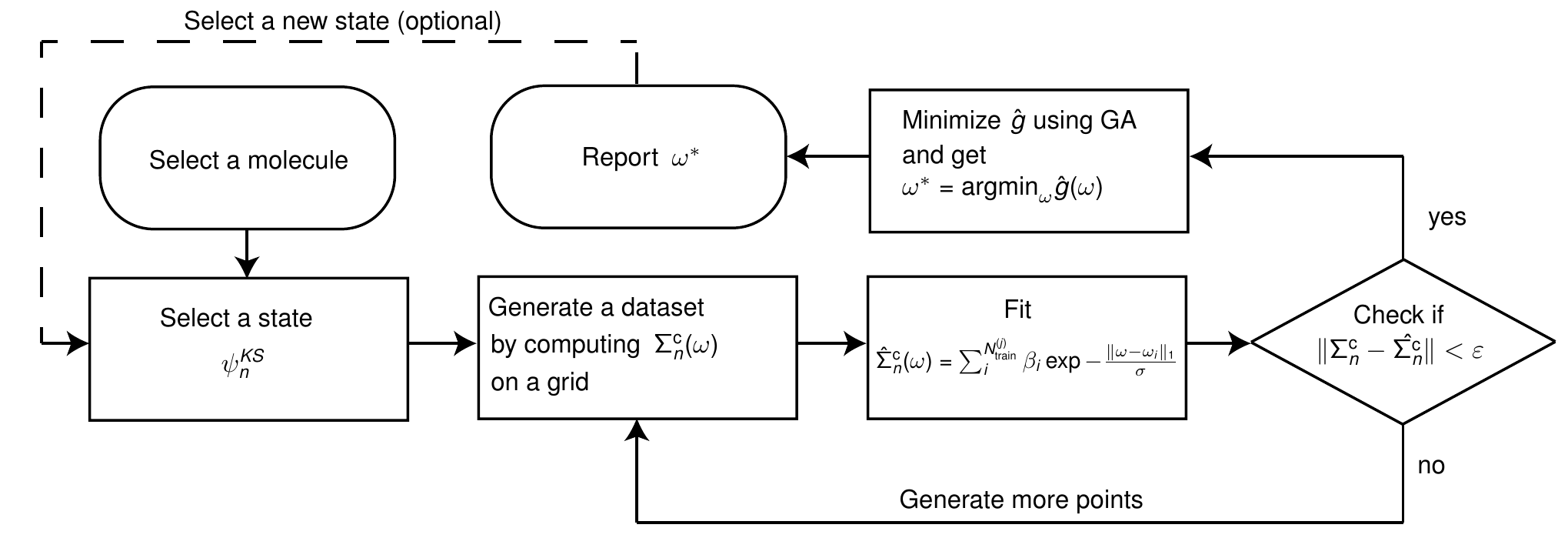}
  \caption{A flow diagram of a machine learning based QP energy computation.}
  \label{fig:flowChart}
\end{figure}

We propose an alternative approach to tackle this problem by combining genetic algorithms (GA) and a KRR-based surrogate model, as summarized in Fig.~\ref{fig:flowChart}. For any QP state $n$, we define an auxiliary fitness function
\begin{equation}
\label{eqn:opt-eqn}
    g_n(\omega) = |(\omega - \omega_{n}) - \Sigma^\mathrm{c}_n(\omega) | + \left|\frac{d{\Sigma^\mathrm{c}_n}(\omega)}{d\omega}\right|.
\end{equation}
The first term on the RHS is minimized by all the solutions of Eq.~(\ref{equ:theory:qpequgw}). In order to exploit the ability of GA of finding a global minimum we enforce the solution with the highest spectral weight to be the global minimum of the fitness function by adding the second term. We can thus determine the global minimum $\omega^*={\text{argmin}}_{\omega}g(\omega)$,
corresponding to the QP energy, by employing a standard elitist genetic-algorithm. 
            
GA is designed to search for an optimal solution in a large search space. This requires several evaluations of the fitness function (self-energy) we cannot afford.
To overcome the limitation of expensive repeated fitness function evaluations in the GA, we adopt an adaptive fitting procedure for $\Sigma^\mathrm{c}_n(\omega)$. 
To this end, we aim at training our kernel-based model with the lowest number of self-energy evaluations. The number of evaluations should be proportionate to the number of poles in the self-energy profile. An almost featureless $\Sigma^\mathrm{c}_n(\omega)$ like the one in Fig.~\ref{fig:selfenergy}(a) should require less evaluation than the one in Fig.~\ref{fig:selfenergy}(b). 
At step $j$ we generate an initial dataset of $N_{\mathrm{D}}^{(j)}$ couples $\{\omega_i,\Sigma^\mathrm{c}_n(\omega_i)\}$ sampled from a coarse grid around $\varepsilon_n^\mathrm{KS}$. The dataset is split in $N_{\mathrm{train}}^{(j)}$ training points and $N_{\mathrm{test}}^{(j)}$ testing points ($N_{\mathrm{D}}^{(j)} = N_{\mathrm{test}}^{(j)} +N_{\mathrm{train}}^{(j)}$). The self-energy is thus approximated by a Laplacian based KRR
\begin{equation}
    \hat{\Sigma}^\mathrm{c}_n(\omega)=\sum_i^{N_{\mathrm{train}}^{(j)}} \beta_i \exp{-\frac{\|\omega-\omega_i\|_1}{\sigma}},
    \label{eqn:laplacian_kernel}
\end{equation}
with the kernel width $\sigma$, trained with 5-fold cross-validation for hyperparameter optimization. Mean Average Error (MAE) between the predicted self-energy and the testing set must be lower than a desired threshold to declare the fitting successful. Otherwise we proceed with step $j+1$ in which $N_{\mathrm{train}}^{(j+1)} = N_{\mathrm{D}}^{(j)}$ and a new set of $N_{\mathrm{test}}^{(j+1)}$ testing points is computed. 
With the fitted analytical expression $\hat{\Sigma}^\mathrm{c}_n(\omega)$ we can compute cost-free the gradient at each frequency point and let the GA explore the continuous landscape of the fitness function.

\section{Results}
The approach has been implemented into VOTCA-XTP~\cite{Wehner2018ElectronicVOTCA-XTP,Tirimbo2020Excited-stateVOTCA-XTP}, a software for calculating the excited-state electronic structure of molecules based on Gaussian orbitals. 
We showcase the $G_0W_0$ HOMO calculation for eight molecules of the $GW$100 dataset~\cite{VanSetten2015GW100:Systems} for which a pole in the self-energy is very close the solution of the QP equation (as in Fig.~\ref{fig:selfenergy}(b)), and which presents the most challenging test for our procedure.

We compare results to a standard grid-search implementation in XTP, which evaluates  $\Sigma^\mathrm{c}_n(\omega)$ at $1001$ grid points to preselect intervals containing solutions to Eq.~(\ref{equ:theory:qpequgw}) followed by a bisection to refine the solutions and finally reporting the one with the highest $Z_n(\omega)$. All the calculations are done with PBE~\cite{Perdew1996GeneralizedSimple} exchange-correlation functionals and with the def2-QZVP basis~\cite{Weigend2005BalancedAccuracy}. For the frequency integration of the self-energy, we employed the Fully Analytic Approach~\cite{Golze2019TheSpectroscopy,Tirimbo2020Excited-stateVOTCA-XTP}. In the KRR, we use $\sigma=1$ Hartree and consider the training converged with a testing MAE below $\varepsilon=0.1$ eV. The GA optimization is performed with a population size of $100$, a mutation probability of $0.1$, an elite ratio of $0.01$ and a crossover probability $0.5$ (uniform). 

\begin{figure}[t]
  \centering
  \includegraphics[width=\linewidth]{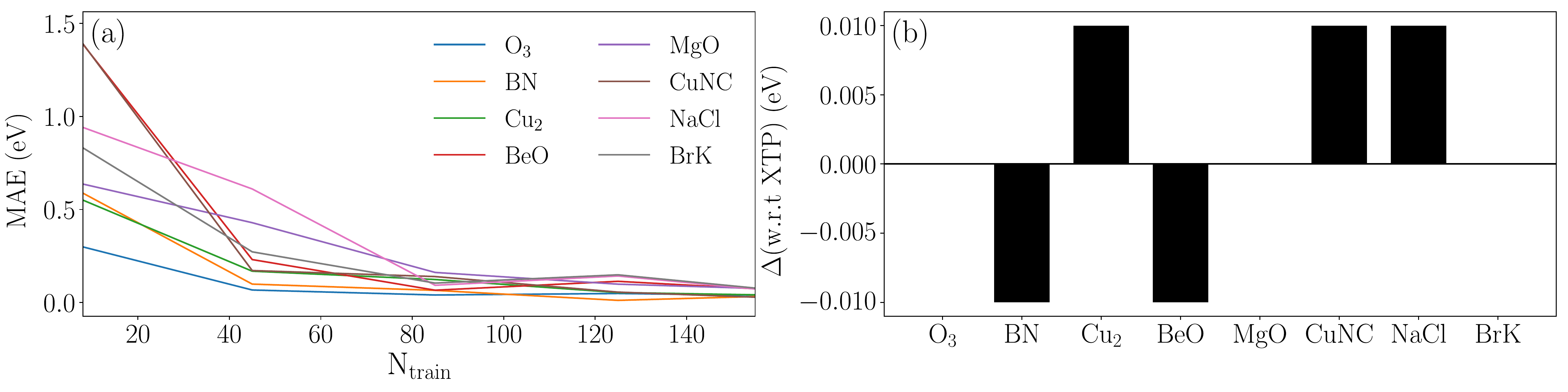}
  \caption{(a) Learning curves for eight $GW$100 molecules of prediction error as a function of training set size. (b) Deviations with respect to the XTP reference calculations of QP energies in eV.}
  \label{fig:lc_bar}
\end{figure}

Fig.~\ref{fig:lc_bar}(a) shows the learning curves of the eight molecules. In most cases the KRR reaches the desired accuracy of $0.1$ eV (MAE on $\Sigma^\mathrm{c}_n(\omega)$) after training with only 150 points. In some cases, like \ch{O_3}, even only approximately 50 points are sufficient. We report in Fig.~\ref{fig:lc_bar}(b) the difference $\Delta$ of the QP energies predicted with the KRR-GA approach with respect to the standard implementation. For all studied molecules, the difference between the two methods is $|\Delta| \leq 0.01$ eV. 

\section{Conclusion and Outlook}
Application to the selection of small molecules with a complicated structure of the self-energy has demonstrated that the surrogate KRR-GA method for the solution of the QP equation is both accurate and computationally efficient compared to standard solution techniques. Its key advantages are the adaptive limitation of the number of expensive self-energy computations and automatic finding of physically relevant solutions. 

The simplicity of the algorithm allows for future expansion and improvements. For example, we will explore the possibility of implementing a multi-population genetic algorithm to search for all the solutions without constraining the one with the highest spectral weight. 

Despite this being a simple ML application we hope to inspire the community to organically include ML methods in their workflow to reduce computational costs, unlocking new possibilities for analyzing larger systems.

\small
\bibliographystyle{unsrt} 
\bibliography{literature.bib} 

\end{document}